\documentclass[aps,prl,twocolumn,10pt,showpacs,superscriptaddress]{revtex4-1}

\usepackage{amsmath}
\usepackage{graphicx}
\usepackage{times}
\usepackage{amssymb}
\usepackage{hyperref}
\usepackage{textcomp}
\usepackage{xcolor}
\usepackage[final]{pdfpages}
\usepackage[T1]{fontenc}

\begin{document}
\title{Signatures of Quantum Coherences in Rydberg Excitons}
\author{P. Gr\"unwald}
\email{Electronic address: peter.gruenwald@uni-rostock.de}
\affiliation{Institut f\"ur Physik, Universit\"at Rostock, 
Albert-Einstein-Strasse 23, D-18059 Rostock, Germany}
\author{M. A{\ss}mann}
\author{J. Heck\"otter}
\author{D. Fr\"ohlich}
\author{M. Bayer}
\affiliation{Experimentelle Physik 2, Technische Universit\"at Dortmund, 
D-44221 Dortmund, Germany}
\author{H. Stolz}
\author{S. Scheel}
\affiliation{Institut f\"ur Physik, Universit\"at Rostock, 
Albert-Einstein-Strasse 23, D-18059 Rostock, Germany}

\begin{abstract}
Coherent optical control of individual particles has been 
demonstrated both for atoms and semiconductor quantum dots. 
Here we demonstrate the emergence of quantum coherent effects in semiconductor Rydberg excitons in bulk Cu$_2$O.
Due to the spectral proximity between two adjacent Rydberg exciton
states, a single-frequency laser may pump both resonances with little dissipation from the detuning. 
As a consequence, additional resonances appear in the absorption spectrum that correspond 
to dressed states consisting of two Rydberg exciton levels coupled to the excitonic vacuum, 
forming a V-type three-level system, but driven only by one laser light source. 
We show that the level of pure dephasing in this system is extremely low. These observations 
are a crucial step towards coherently controlled quantum technologies 
in a bulk semiconductor.
\end{abstract}

\maketitle

\paragraph{Introduction.}
Rydberg atoms are understood to be those whose valence electron has been 
promoted to a quantum state with principal quantum number $n\gg1$. These exotic 
quantum states feature large coupling strengths to other Rydberg states and 
extremely long life times (scaling with $n^2$ and $n^{3\cdots4.5}$, 
respectively), making them an excellent choice for the study of quantum 
aspects of light-matter interaction \cite{Mesched85}. Rydberg atoms have been 
used in the first direct observation of quantum collapses \cite{Rempe87} as 
well 
as quantum non-demolition experiments \cite{Haroche01}. In recent years, they 
have matured into a promising platform for quantum information research 
\cite{Saffman}.

Rydberg physics is not confined to the realm of atomic physics.
Recently, Rydberg states with principal quantum numbers up to $n=25$ 
were observed in the semiconductor $\mathrm{Cu_2O}$ 
\cite{Kazimierczuk2014}. Here, Mott-Wannier excitons, bound states of 
electrons and holes, substitute for real atoms. A number of concepts 
developed within the framework of Rydberg atoms are displayed by these exotic 
states, such as typical scaling laws for the natural linewidth or the 
oscillator strength ($\propto n^{-3}$ \cite{HaugKoch,Elliott61}). Additionally, 
a quantum defect associated with an effective
quantum number $n^*=n-\delta_\ell$ \cite{Rydberg89} was observed and shown 
to be mainly due to the nonparabolicity of the valence bands 
\cite{Schoene2015,Schoene2016}.

\begin{figure}[htbp]
\includegraphics[width=7cm]{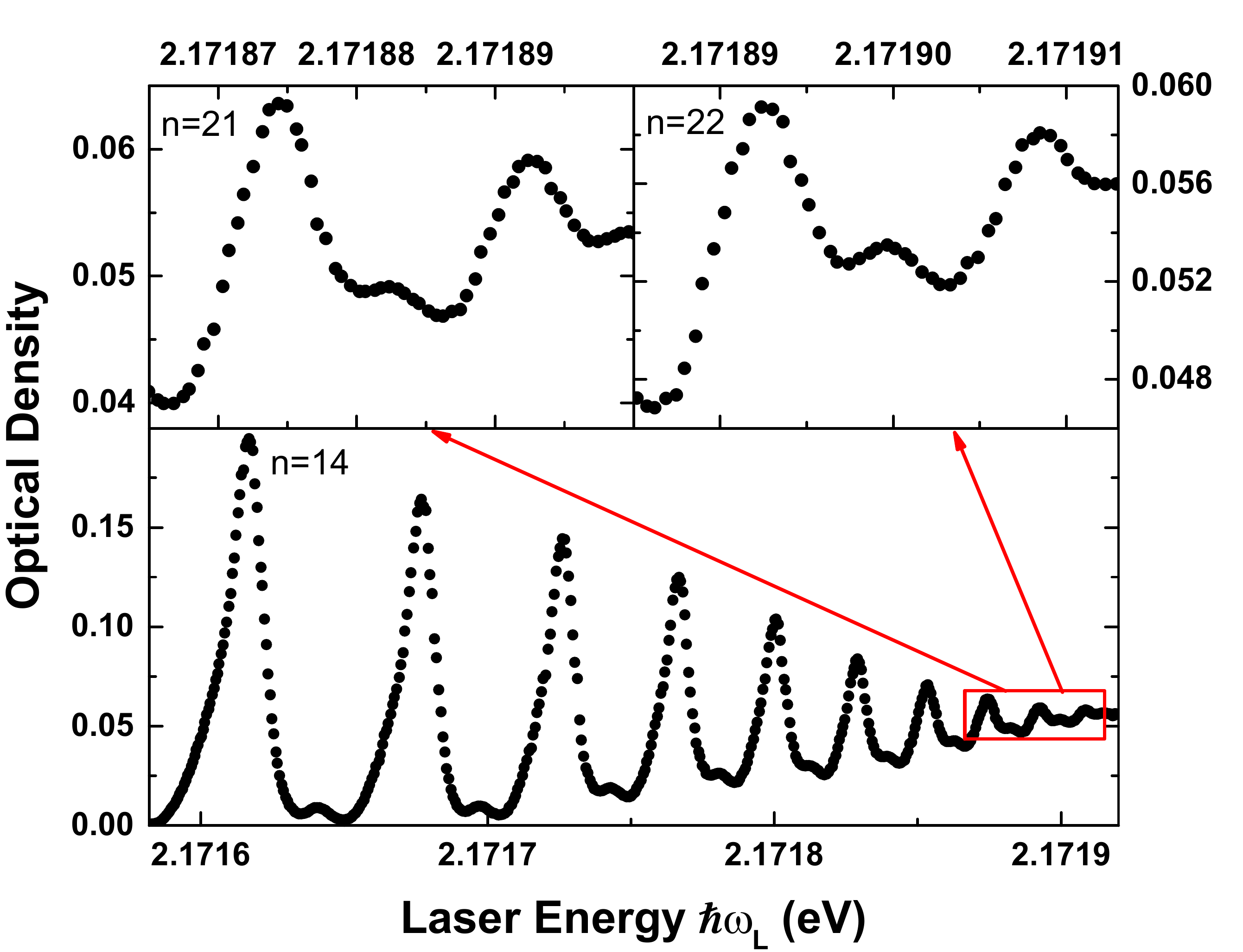}
\caption{Experimental absorption spectra showing additional resonances between 
isolated Rydberg exciton resonances.}
\label{fig.ExpAdd}
\end{figure}

A hitherto unexplained feature is the appearance of additional resonances in the 
absorption spectra of Rydberg excitons for principal quantum numbers larger than $n\gtrsim 12$ (see Fig. \ref{fig.ExpAdd}). They occur 
halfway between each pair of excitonic resonances after excitation with a 
narrow-bandwidth laser, but vanish for large laser powers, where 
Rydberg blockade sets in. These resonances constitute a significant deviation 
from simple Rydberg models, as well as established methods of many-body 
physics~\cite{Baldereschi73, Baldereschi74,Thewes15,Schoene2015,Schoene2016}. We 
will demonstrate here that they represent coherent 
(quantum) interactions between exciton states of different principal quantum 
numbers.

In the following, we consider quantum coherence in terms of the off-diagonal 
elements of the density matrix in Fock basis. An effect is based on quantum coherence
if these off-diagonal elements are a necessary constituent of said effect. 
For open quantum systems, and in particular for the emission of light from a 
given source, this is achieved using a dominant coherent drive of the source of the 
emission. Such a dominance is usually diminished by dissipative effects such as 
energy and phase relaxation (dephasing). Semiconductor systems are typically 
prone to strong phase relaxations due to phonon coupling and other many-body 
effects. Nevertheless, in low-dimensional systems such as quantum dots in microcavities, quantum coherent processes 
have been observed \cite{QDstrong1,QDstrong2,Shih-1}. However, so far no such 
coherent effects could be demonstrated in bulk systems such as those used for 
Rydberg-exciton generation.

The quantitative description of absorption and emission processes in 
semiconductors generally requires both the medium properties as well as a 
quantum-optical description of the emitters, here the excitons. For 
example, the fluorescent emission of a coherently pumped quantum well required 
the combination of both the absorption (not to be confused with the absorption 
spectrum) as well as the spectral intensity distribution of the excitons 
\cite{QWell13,QWell16}. In that case, the necessity to combine both processes follows 
from Kirchhoff's law \cite{Henneb1} and from the quantum-optical input-output 
relations \cite{Dima1}. More recently, such an approach has also been applied 
to highly doped quantum wells in thermal pumping \cite{Vasanelli15}. Using a similar approach, it 
becomes possible to infer that quantum coherence has been achieved in Rydberg exciton systems
directly from the experimental absorption spectra.

\paragraph{Absorption Spectra.} 
The absorption spectrum of a semiconductor is commonly measured via a frequency-
insensitive detector such as a photodiode placed behind the sample. 
This device registers the spectrally integrated transmission of 
light, which is then compared to the full laser intensity 
impinging on the crystal. A spectral distribution 
as shown in Fig.~\ref{fig.ExpAdd} is obtained as a function of the laser frequency
 $\omega_\text L$.  The lost intensity is often assumed to be equal to the 
absorption, thereby neglecting secondary emission and losses due to scattering. 
The commonly applied model for the absorption spectrum $a(\omega)$ 
of an exciton with principal quantum number $n$ follows the Toyozawa formula
\cite{Toyozawa64}
\begin{equation}
a(\omega)=C_n\frac{\tfrac{\Gamma_n}{2}-2q_n(\omega_n-\omega)}{(\tfrac{\Gamma_n}{
2})^2+(\omega_n-\omega)^2},\label{eq.Toyo}
\end{equation}
where $C_n$ is proportional to the oscillator strength and $\omega_n$ and $q_n$ 
are the resonance frequency of the exciton and the asymmetry parameter of the 
line shape, respectively. The natural linewidth $\Gamma_n$ should scale 
with $n^{-3}$, as known from atomic systems. According to Elliott 
\cite{Elliott61}, in $\mathrm{Cu_2O}$ one also observes $C_n\propto n^{-3}$.

However, Eq.~(\ref{eq.Toyo}) only describes the probability density of absorbing a 
photon at frequency $\omega$ by the crystal, whereas the measurement setup
detects the transmitted light as well as contributions from resonance fluorescence, which includes light scattered or reemitted along the forward direction. Therefore, in the event that a photon is 
indeed absorbed, $a(\omega)$ is scaled by the resonance fluorescence contributions from the excitons $I_\text X(\omega,\omega_\text L)$~\cite{Malerba11}. That 
scaled spectrum impinges on the frequency-insensitive detector yielding a spectrum 
\begin{equation}
  A(\omega_\text L)=\int\limits_{-\infty}^{\infty} d\omega\, a(\omega) I_\text X(\omega,\omega_\text L).
  \label{eq.Ameas}
\end{equation}
The determined ``absorption'' spectrum of the $\mathrm{Cu_2O}$ crystal thus also 
contains components from coherent and incoherent emissions of the sample. Here the coherent contributions correspond to phase-preserving elastic scattering processes, which replicate the excitation field. Meanwhile, the incoherent contributions include inelastically scattered fields, which can result, for example, in the appearance of the Mollow triplet for large Rabi frequencies \cite{Matthiesen12}. 

$I_\text X(\omega,\omega_\text L)$ is determined by the spectral intensity distribution of the excitons $S_\text X(\omega,\omega_\text L)$ via
\begin{equation}
  I_\text X(\omega,\omega_\text L)=g^2S_\text X(\omega,\omega_\text L),
\end{equation}
with $g$ being proportional to the dipole transition moment $d$ between the excitonic 
and vacuum states. In turn, $S_\text X(\omega,\omega_\text L)$ is directly connected to the 
quantum optical dynamics of the exciton states driven by the laser beam. Modelling
adequately the quantum states involved in excitation and relaxation processes
hence suffices to determine $A(\omega_\text L)$ and compare with the experiments.

\paragraph{Master Equation Approach.} 
In order to obtain the correct exciton dynamics, we employ the following 
description. A single exciton resonance is modeled by only two 
states, the excited state $|n\rangle$  and the excitonic vacuum $|0\rangle$ as 
the ground state. Both states are dipole-coupled (dipole moment 
$d_n$) by the pump laser with Rabi frequency $\Omega_n$. Further, the excited 
state undergoes radiative, Markovian damping with energy relaxation rate 
$\Gamma_n$. According to standard Wigner--Wei\ss{}kopf theory for the damping, 
we find that 
\begin{equation}
\Gamma_n\propto d_n^2\propto n^{-3},\quad\Omega_n\propto d_n\propto 
n^{-3/2}.\label{eq.deps}
\end{equation}
The Hamiltonian and the corresponding master equation for the density operator 
follow a Bloch theory:
\begin{equation}\label{eq.Ham_n}
\begin{split} 
\hat H_n=&\hbar\left[\delta_{n}\hat A_{n,n}+\Omega_n(\hat A_{0,n}+\hat 
A_{n,0})\right],\\
\frac{d\hat\varrho_n}{dt}=&\frac{1}{i\hbar}[\hat H_n,\hat\varrho_n]
+\frac{\Gamma_n}{2}\mathcal{L}_{\hat A_{0,n}}[\hat\varrho_n],\\
\mathcal{L}_{\hat X}[\hat\varrho]=&2\hat X\hat\varrho\hat X^\dagger
-\hat X^\dagger\hat X\hat\varrho-\hat\varrho\hat X^\dagger\hat X,
\end{split}
\end{equation}
with $\hat A_{j,k}=|j\rangle\langle k|$ being the excitonic operators and $\delta_n=\omega_n-\omega_\text L$.
Here, $\mathcal L_{\hat X}$ describes the Lindblad term for operator $\hat X$, 
in this case the exciton annihilation operator $\hat A_{0,n}$.

It is instructive to discuss the limit of weak light-matter coupling first, as
that is usually expected in these experiments.
For low excitation, $\langle\hat A_{n,n}\rangle\ll1$, 
the incoherent spectral components from the excitons can be neglected as well as
excitation-induced dephasing via exciton-exciton scattering. 
Thus, the spectral distribution of the excitons reduces to a Rayleigh 
component \cite{WelVo} $S_\text X(\omega,\omega_\text L)=|\mathcal C|^2\delta(\omega-\omega_\text 
L)$, where $|\mathcal C|^2$ is the coherent part of the exciton intensity. Thus,
we find in this limit
\begin{equation}
 A(\omega_\text L)=g^2|\mathcal C|^2a(\omega_\text L), 
\end{equation}
meaning that fitting the observed spectra in Ref.~\cite{Kazimierczuk2014} to Eq.~(\ref{eq.Toyo})
is sufficiently accurate if $g^2|\mathcal C|^2$ can be regarded as insensitive to 
variations of the laser frequency and the principal quantum number $n$. For higher laser 
powers with respect to the light-matter couplings, this condition becomes invalid and $|\mathcal C|^2$ must be investigated in more detail.
In the steady-state solution of Eq.~(\ref{eq.Ham_n}), $|\mathcal C|^2=|\langle\hat A_{n,0}\rangle|^2\approx\langle\hat A_{n,n}\rangle$,
and the measured absorption spectrum becomes
\begin{equation}
A(\omega_\text 
L)=g^2\Omega_n^2C_n\frac{\tfrac{\Gamma_n}{2} - 2 q_n\delta_n} 
{\left[(\tfrac{\Gamma_n}{2})^2+2\Omega_n^2+\delta_n^2\right]^2}.
\label{eq.cohfull}
\end{equation}

Comparing Eq. (\ref{eq.cohfull}) with Eq.~(\ref{eq.Toyo}), a 
few results should be noted. 
First, the oscillator strength is calculated by integrating an 
absorption peak over all laser frequencies $\omega_L$. In the linear regime, where
$\Omega_n\ll\Gamma_n$, all $n$-dependencies apart from $C_n$ cancel out in the 
integrated transmission. Thus, the expected $n^{-3}$-dependence of the 
oscillator strength is precisely what is observed in the experiment.
Second, the corrected line shape of the spectrum differs slightly 
from the Toyozawa formula, with the denominator being squared. This yields 
deviations to the fit parameters $q_n$ and $\Gamma_n$ in a least-squares fit.  
The line shape of isolated resonances 
can be fitted well to both models, Eqs.~(\ref{eq.Toyo}) and 
(\ref{eq.cohfull}), making a differentiation impossible (see SOM). This means that our corrected model can fully explain the standard Toyozawa line shape observed in earlier experiments.
Overall we can conclude that, in the linear regime and considering only 
isolated resonances, we find no indication of additional resonances.

Third, and notwithstanding the previous statements the approximation of using 
solely Toyozawa's formula breaks down, however, when reaching 
the nonlinear regime by increasing either the laser intensity or the principal 
quantum number $n$ of the exciton resonance. For $\Omega_n\lesssim\Gamma_n$, 
power broadening sets in, yielding a perceived larger line width and a steeper slope 
for the fitted oscillator strength. Given the scaling $\Omega_n/\Gamma_n\propto n^{3/2}$ 
from Eq.~(\ref{eq.deps}), in the strong-coupling limit ($\Omega_n\gg\Gamma_n$)
the area under the absorption peak scales as $n^{-15/2}$ compared to the $n^{-3}$-scaling 
in the weak-coupling regime. 

\paragraph{Additional resonances in the absorption spectrum.}
At intermediate excitation strengths $\Omega_n\lesssim\Gamma_n$, the incoherent 
part of the exciton spectrum becomes relevant, i.e. $S_\text X(\omega,\omega_\text L)$ is no longer
proportional to a $\delta$-distribution but instead broadens. The incoherent 
spectrum of a single two-level system and its quantum features are well documented in 
the literature \cite{Mollow69,Schuda74}. One may assume that the additional
resonances result from strong-coupling effects, representing light-matter dressed 
states. However, at the given laser intensities, we are not yet in the 
strong-coupling regime, and thus, only a single peak occurs in the exciton 
spectrum for one isolated Rydberg resonance. Additional peaks
resulting from strong coupling effects would show different spectral behaviour, 
such as the Mollow-triplet.
Hence, an isolated exciton state $n$ is insufficient to explain the additional 
resonances. 

\begin{figure}[htbp]
\includegraphics[width=7cm]{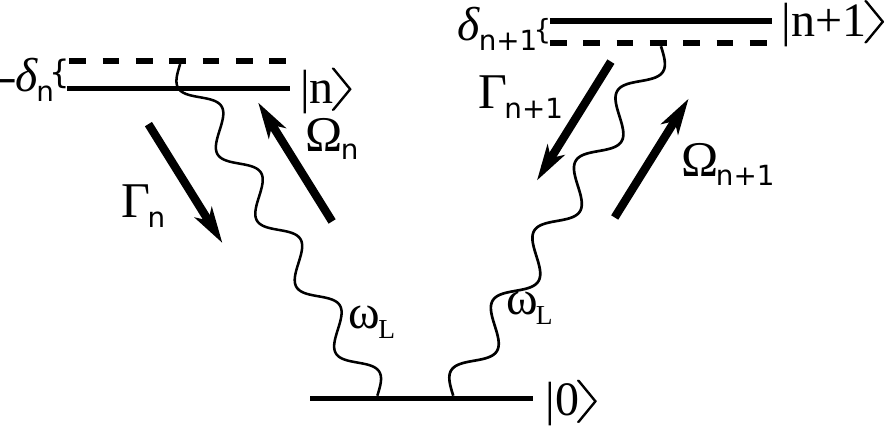}
\caption{Model for two Rydberg excitons with principal quantum 
numbers $n$ and $n+1$, coupled to the excitonic vacuum state 
$|0\rangle$.}
\label{fig.3LS}
\end{figure}

However, excitons with large $n$ can no longer be considered spectrally isolated, 
see Fig.~\ref{fig.ExpAdd}. These resonances are spectrally close, e.g., 
$|\omega_{15}-\omega_{16}|\approx 49~\mu\text{eV}/\hbar=12~\text{GHz}.$
Hence, tuning the laser frequency around a high-$n$ resonance, different states $n, n+1$ 
of the exciton can be populated with limited influence by the detuning.
It is therefore appropriate to consider a system of two exciton states $|n\rangle$ and 
$|n+1\rangle$ as the possible excited states, and the excitonic vacuum 
$|0\rangle$ as their common ground state. The resonances do not couple directly 
to one another but they are dipole-coupled to the vacuum state and driven by the same laser 
light, yielding a V-type three-level system with a single pump laser of frequency 
$\omega_\text L$ (see Fig.~\ref{fig.3LS}). Note that we do not consider two excitons, but the presence of a single exciton in a system with different possible excited quantum states.

Comparing with Eq.~(\ref{eq.Ham_n}), we now take into account the sum of the two 
Hamiltonians, $\hat{H}_n$ and $\hat{H}_{n+1}$, with the corresponding Lindblad 
terms and relaxation rates $\Gamma_n$ and $\Gamma_{n+1}$:
\begin{equation}\label{eq.Ham_3LS}
\begin{split} 
  \hat H_\text{3LS}=&\hat H_n+\hat H_{n+1},\\
  \frac{d\hat\varrho_\text{3LS}}{dt}=&\frac{1}{i\hbar}[\hat H_\text{3LS},
  \hat\varrho_\text{3LS}]\\
  &+\frac{\Gamma_n}{2}\mathcal{L}_{\hat A_{0,n}}[\hat\varrho_\text{3LS}]+
  \frac{\Gamma_{n+1}}{2}\mathcal{L}_{\hat A_{0,n+1}}[\hat\varrho_\text{3LS}].
\end{split}
\end{equation}
The common ground state $|0\rangle$ permits a transfer of an excitation from one 
exciton state to the other, forming an effective quadrupolar coupling, i.e. $\langle\hat 
A_{n,n+1}\rangle\neq 0$, see SOM for details. Hence, the coherent drive induces an effective transition between the exciton states, 
whose transition rate is directly linked to the Rabi frequencies $\Omega_n$ and $\Omega_{n+1}$.

\begin{figure}[htbp]
\includegraphics[width=\linewidth]{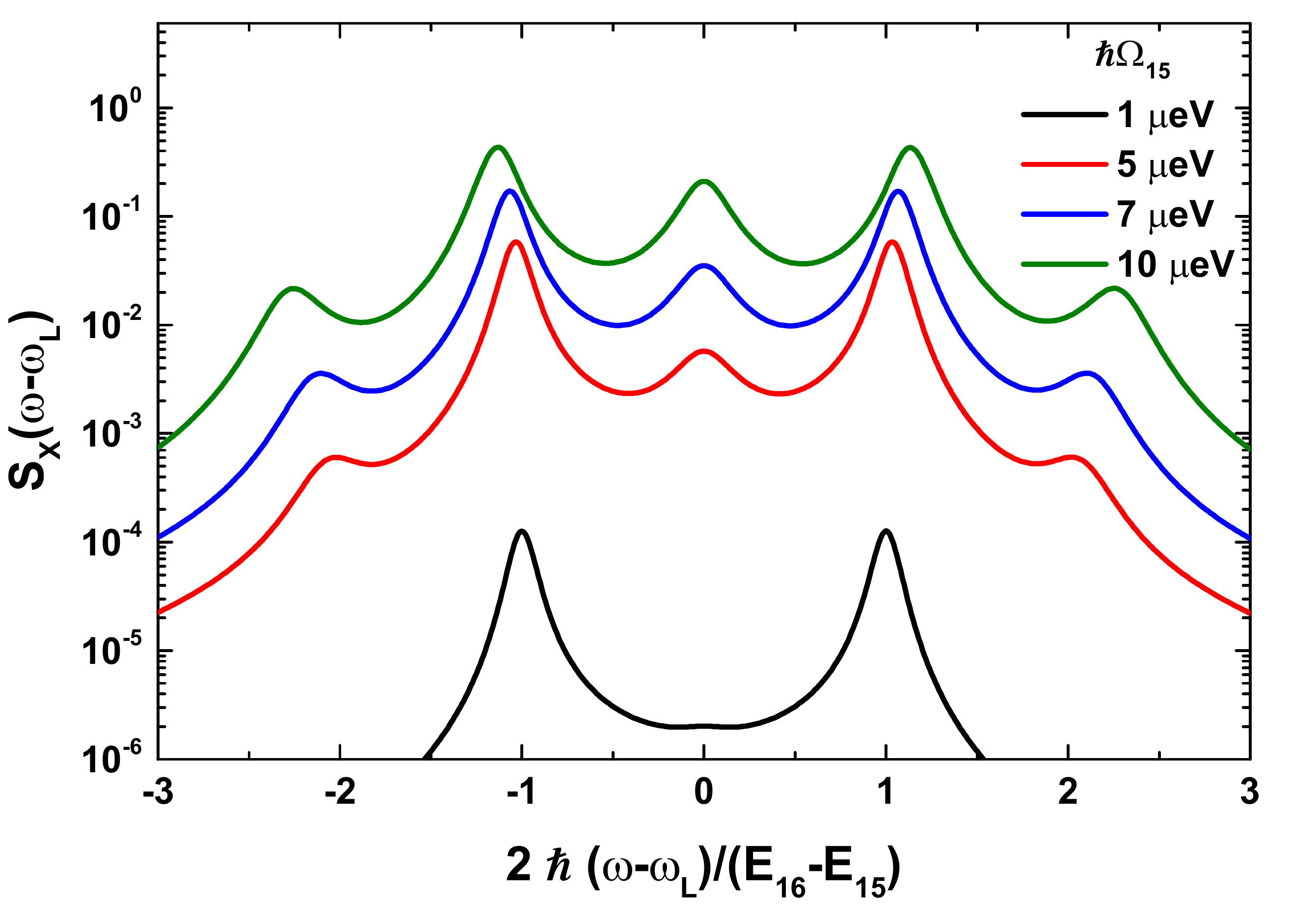}
\caption{Incoherent exciton-intensity spectrum $S_\text X(\omega)$ for two resonances with $n=15,16$, 
coupled to the excitonic vacuum with the parameters taken from 
Ref.~\cite{Kazimierczuk2014}. The laser is tuned to the center between the 
resonances and the Rabi frequencies are
$\Omega_{16}=\Omega_{15}(\tfrac{15}{16})^{3/2}$.}
\label{fig.3LS_X_spec}
\end{figure}

\begin{figure}[htbp]
\includegraphics[width=\linewidth]{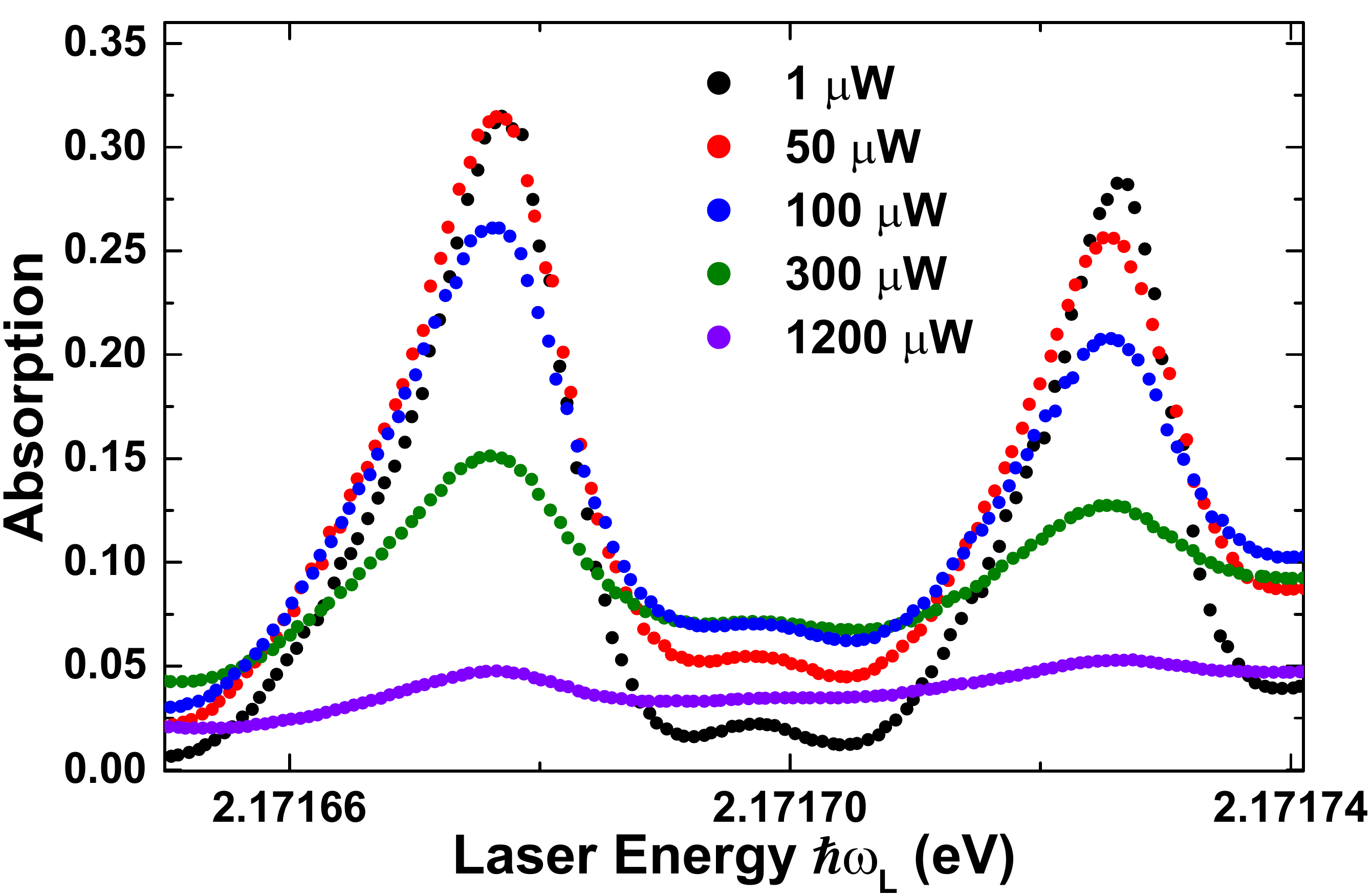}
\includegraphics[width=\linewidth]{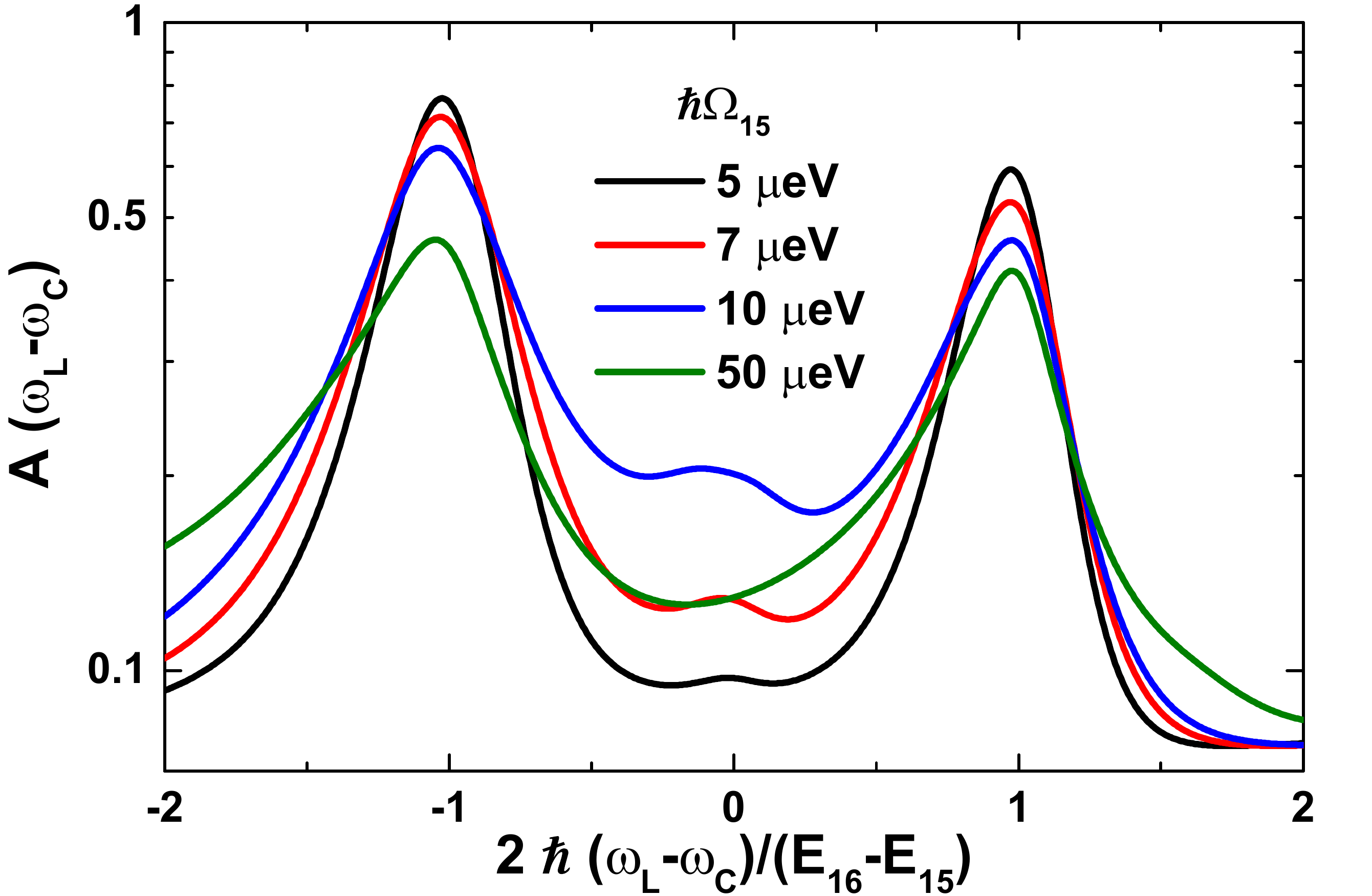}
\caption{Top: experimental absorption spectra for $n=15,16$ and different laser 
intensities as shown in the legend. Bottom: absorption spectrum for the two 
resonances $n=15,16$ for different Rabi frequencies with 
$\Omega_{16}=\Omega_{15}(\tfrac{15}{16})^{3/2}$.}
\label{fig.Rabivar}
\end{figure}

The incoherent part of the three-level exciton intensity spectrum $S_\text X(\omega,\omega_\text L)$ is displayed in Fig.~\ref{fig.3LS_X_spec} for the laser frequency $\omega_\text L=\tfrac{\omega_{15}+\omega_{16}}{2}$. 
With increasing Rabi frequency (curves from bottom to top), an internal 
resonance builds up. This resonance corresponds to a dressed state of the two 
excitons, formed by the effective quadrupole coupling.
The measured absorption spectrum $A(\omega_\text L)$, following Eq.~(\ref{eq.Ameas}), now
includes $S_\text X(\omega,\omega_\text L)$ from the three-level Hamiltonian and 
the Toyozawa formula (\ref{eq.Toyo}). It features prominently the additional resonances 
from the experiments (see bottom panel in Fig.~\ref{fig.Rabivar}).

To account for the discrepancies between system parameters determined via fitting
Toyozawa's formula and the complex three-level model, we apply the following rules: the decay
rates $\Gamma_n$ follow the basic $n^{-3}$ law, fitted to the lower $n$-
resonances. Deviations from this behaviour are thus thought to be included in the full
description. The $q_n$ are taken from the fit parameters. For a simple model discussion
on their behaviour see the SOM. The Rabi frequencies were not 
fitted in~\cite{Kazimierczuk2014}, so we will use values that reproduce the experiments.

Further increase of the laser intensity results in two effects (see top panel 
in Fig.~\ref{fig.Rabivar}). First, the additional resonances develop an
asymmetry towards the lower-frequency side of the spectrum. Second, for even 
higher field intensities, they start to disappear, while the absorption in 
general decreases. The former effect follows directly from our model 
Hamiltonian, where the combination of symmetric exciton intensity spectrum and 
the asymmetric Toyozawa formula tilts the additional resonances (see lower 
panel in Fig.~\ref{fig.Rabivar}). The regime of strong pumping is signified by 
the latter effect. The dressed states are shifted out of the spectral region
by the large Rabi frequency. Furthermore, for strong driving,
the dipole- and quadrupole moments decrease because of the 
saturated level occupation. In that case, the effective system reverts back to the 
isolated resonance scenario. Also, at this point the Rydberg blockade 
sets in \cite{Kazimierczuk2014}. If one exciton is excited to a Rydberg state, 
the excitation of another one becomes suppressed.  Note that Rydberg blockade depends strongly on 
parameters like sample thickness and the exact geometry used for excitation and 
has thus not been included in the theoretical results shown in the lower panel of 
Fig.~\ref{fig.Rabivar}. 

\begin{figure}[htbp]
\includegraphics[width=\linewidth]{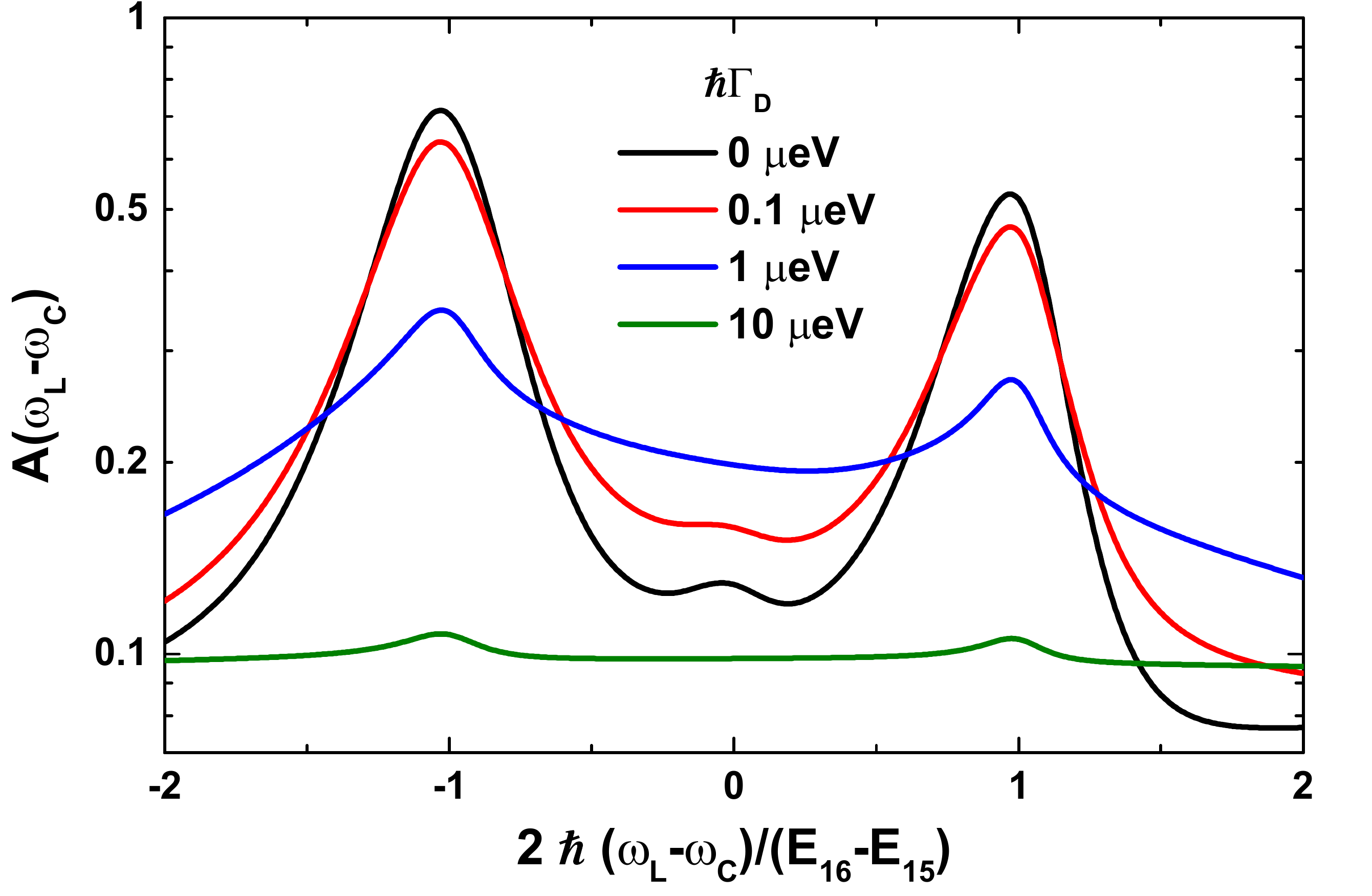}
\caption{Absorption spectrum for the two 
resonances $n=15,16$ for different dephasing rates $\Gamma_\text D$ and 
$\hbar\Omega_{15}=7$ $\mu$eV and $\Omega_{16}=\Omega_{15}(\tfrac{15}{16})^{3/2}$.}
\label{fig.Deph}
\end{figure}

So far, we have limited our discussion to energy relaxation, neglecting the 
influence of pure dephasing. If the additional resonances are indeed based on 
quantum coherent coupling of the different resonances, increasing the pure dephasing 
rate (by inclusion of another Lindblad term $\mathcal L_{\hat A_{n,n}}[\hat\varrho]$ 
into the master equation) should influence their magnitude. Indeed, as shown in 
Fig.~\ref{fig.Deph}, already for dephasing rates close to the pumping 
rate, the resonances disappear. Two conclusions can be drawn from 
this. First, radiationless dephasing is surprisingly small in Rydberg 
excitons in $\mathrm{Cu_2O}$ and second, the additional resonances appearing 
between the exciton resonances are clear signatures of quantum coherence between adjacent
exciton states. Remarkably, these coherences between exciton states with 
consecutive quantum numbers are driven by the incoherent part of the exciton spectrum 
instead of the coherent part, which drives the coherences between 
the ground state and the exciton states.

We also note that the curves for higher dephasing rates are very similar to the experimental 
curves for higher pumping, see top panel in Fig.~\ref{fig.Rabivar}. It seems reasonable that, when
increasing the laser intensity beyond saturation, phonons become dominantly excited and induce
stronger dephasing rates. This can only be better understood by a thorough analysis of the 
decay rates and Rabi frequencies present in the experiments.

\paragraph{Summary and Conclusions.} We have analyzed the quantum coherent 
properties of the Rydberg exciton spectrum in $\mathrm{Cu_2O}$. The 
observed absorption in different experimental situations includes, besides the 
absorption-induced attenuation, also the intensity spectrum of the Rydberg 
excitons. In the linear regime of a coherently driven system, this additional 
effect simply scales the absorption spectrum. For stronger excitation, the 
incoherent part of the exciton spectrum induces additional resonances, which 
were also observed in the experiments. Their appearance indicates the onset of 
measurable quantum coherence between exciton states of consecutive principal 
quantum number, as well as very low levels of pure dephasing. 

Our model for the absorption spectrum of Rydberg excitons
accounts for a number of hitherto unexplained features, which should also be 
applicable to a variety of similar semiconductor experiments. Rydberg excitons 
in $\mathrm{Cu_2O}$ have now reached a stage at which coherent quantum 
effects become visible, and the precise nature of the excitonic Hamiltonian 
becomes detectable. This allows for controlled quantum manipulation and state 
generation for excitons and opens up the possibility for developing 
semiconductor quantum technologies.

\paragraph{Acknowledgements.} We gratefully acknowledge support by the 
Collaborative Research Centre SFB 652/3 'Strong correlations in the radiation 
field' and the International Collaborative Research Centre TRR 160 'Coherent 
manipulation of interacting spin excitations in tailored semiconductors', both 
funded by the Deutsche Forschungsgemeinschaft.

\section*{Supplemental Material}

\subsection*{Comparison of fitted Absorption Spectra}

The experimental spectra from~\cite{Kazimierczuk2014}
were fitted to the Toyozawa formula instead of the absorption sepctrum for low
driving strength, from here on denoted as Rabi spectrum, cf.~Eq.~(1) and (7)
from the main text, respectively. Both functions may be easily fitted to the experiments. To 
compare them theoretically, we will use the following parameters: the apparent resonance
as the maximum of the distribution, the FWHM value, and the peak area.

The maximum of the distribution is given for a certain detuning between laser and actual exciton resonance
$\delta_n=\omega_n-\omega_L\neq0$ due to the asymmetry. Hence, for
$q_n\to0$, $\delta_n$ should become 0 as well for either model function. With this
in mind, we can determine the maximum unambigously to be
\begin{equation}
  \delta_{n,\text{max}}=\frac{\Gamma_n}{4q_n}\left[1-\sqrt{1+4q_n^2}\right]
\end{equation}
for the Toyozawa formula and 
\begin{equation}
  \delta_{n,\text{max}}=\frac{\Gamma_n}{6q_n}\left[1-\sqrt{1+3q_n^2\left(1+\tfrac{8\Omega_n^2}{\Gamma_n^2}\right)}\right]
\end{equation}
for the Rabi spectrum. In case of $\Omega_n\ll\Gamma_n$, we find a different relation between $\Gamma_n$ and $q_n$, but without a theory
for $q_n$ this can easily be included in the fit parameters. Moreover, in the region where $8\Omega_n^2/\Gamma_n^2\approx1$ and for $q_n\ll1$,
the maxima for both models become 
\begin{equation}
  \delta_{n,max}\approx\frac{\Gamma_nq_n}{2}.
\end{equation}

The FWHM value follows, after some lengthy algebra as
\begin{equation}
  \text{FHWM}_n=\sqrt{3+\frac{1-\sqrt{1+4q_n^2}}{q_n^2}}\Gamma_n
\end{equation}
for the Toyozawa formula. The Rabi spectrum FWHM value requires the solution of a fourth-order polynomial. 
More interestingly, when neglecting the asymmetry, we find for the Rabi spectral width
\begin{equation}
	\text{FHWM}_n=\sqrt{\sqrt2-1}\sqrt{\Gamma_n^2+8\Omega_n^2}.
\end{equation}
Power broadening arises for sufficiently large $n$, whereas the overall width is scaled down by a factor of around 1.55. Crucially, the $n$-dependence of the linewidth for $\Gamma_n\gg\Omega_n$ is not changed as the scaling is constant. For a similar small change of the width for $q_n\neq0$ for both 
model functions, we again cannot detect differences between the models for sufficiently low ratios of $\Omega_n/\Gamma_n$.

\begin{figure}[htbp]
\includegraphics[width=\linewidth]{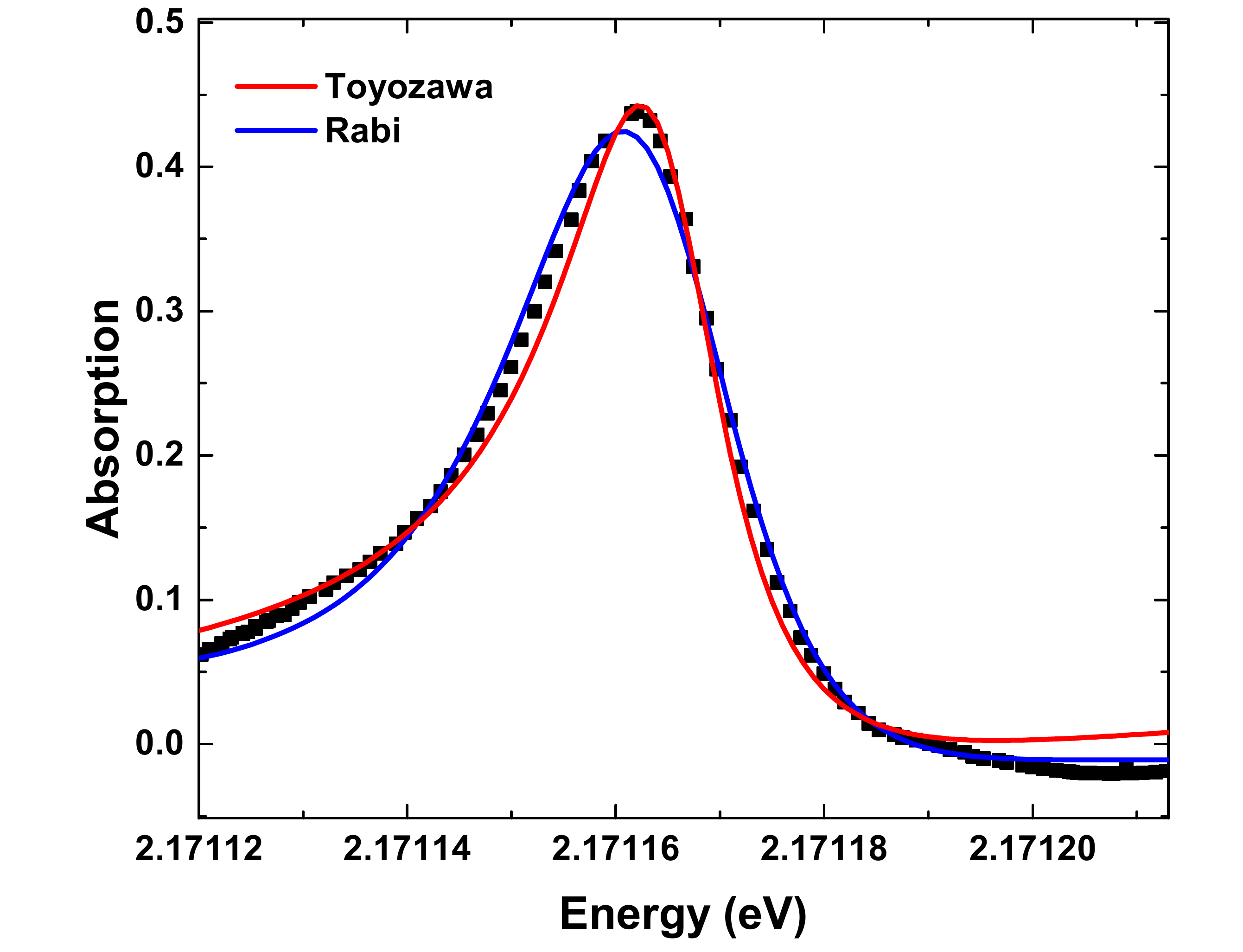}
\caption{Fitted absorption spectrum of the exciton resonance $n=10$, with the Toyozawa and the Rabi-model. }
\label{fig.comp}
\end{figure}

Finally, the integrated peak area $F_n$ reads as
\begin{equation}
	F_n=\pi C_n
\end{equation}
for the Toyozawa formula and
\begin{equation}
	F_n=\frac{\pi}{4} \frac{\Gamma_n g_n^2 \Omega_n^2}{\left(\left(\frac{\Gamma_n}{2}\right)^2+2 \Omega_n^2\right)^{3/2}} C_n
\end{equation}
for the Rabi model. Inserting the known dependencies of the different parameters we find that both scale only with $C_n$ for low pumping strengths, while for larger ones, the Rabi area would scale with $F_n\propto n^{-9/2}C_n$. Overall, we thus deduce that for low pumping strengths no detectable difference between these models can be found by freely fitting the model parameters to the experiments. Exemplarily, we have fitted the resonance $n=10$ in Fig.~\ref{fig.comp} with both models.

\subsection*{Dynamics of Three-Level-System}
Applying the master equation~\cite{WelVo} for the three-level system, Eq.~(8) from the main text,
we can derive the equations of motion for the relevant occupations $\langle A_{jj}\rangle$
and coherences $\langle\hat A_{0j}\rangle$, $j=n,n+1$, respectively, which read as
\begin{align}
  \frac{d}{dt}\langle A_{jj}\rangle=&-\Gamma_j\langle\hat A_{jj}\rangle-i\Omega_j(\langle\hat A_{j0}\rangle
  -\langle\hat A_{0j}\rangle),\label{eq.Ajj}\\
  \frac{d}{dt}\langle A_{0j}\rangle=&-(i\delta_j+\tfrac{\Gamma_j}{2})\langle\hat A_{0j}\rangle-i\Omega_j(\langle\hat A_{00}\rangle
  -\langle\hat A_{jj}\rangle)\nonumber\\
   &+ i\Omega_{k}\langle\hat A_{kj}\rangle.\label{eq.A0j}
\end{align}
Here, $k=(n+1,n)$ for $j=(n,n+1)$. Equation~(\ref{eq.Ajj}) and the first line of 
Eq.~(\ref{eq.A0j}) represent the terms known from an isolated exciton resonance
based on Eq.~(5) from the main text. The only difference is due to the adjusted completeness relation
\begin{equation}
  \hat 1=\hat A_{00}+\hat A_{nn}+\hat A_{n+1\,n+1}.
\end{equation}
However, the second line in Eq.~(\ref{eq.A0j}) stems from the laser driven resonance $k$
and couples the two excited states. It is based on the common ground state of both resonances allowing
an effective quadrupolar coupling via the intermediate state $|0\rangle$.
The equation of motion for such a coupling $\langle\hat A_{kj}\rangle$ then reads as
\begin{align}
  \frac{d}{dt}\langle\hat A_{kj}\rangle=&-[i(\delta_j-\delta_k)+\tfrac{\Gamma_j+\Gamma_k}{2}]\langle\hat A_{kj}\rangle\nonumber\\
  &-i\Omega_j\langle[\hat A_{kj},\hat A_{j0}]\rangle-i\Omega_k\langle[\hat A_{kj},\hat A_{0k}]\rangle\\
  =&-[i(\delta_j-\delta_k)+\tfrac{\Gamma_j+\Gamma_k}{2}]\langle\hat A_{kj}\rangle\nonumber\\
  &-i\Omega_j\langle\hat A_{k0}\rangle+i\Omega_k\langle\hat A_{0j}\rangle.
\end{align}
From the above, two effects couple the resonances $n$ and $n+1$, the completeness relation and the nonvanishing value of 
$\langle\hat A_{n,n+1}\rangle$. 

\subsection*{Model for asymmetry parameter}
\begin{figure}[h]
  \includegraphics[width=7cm]{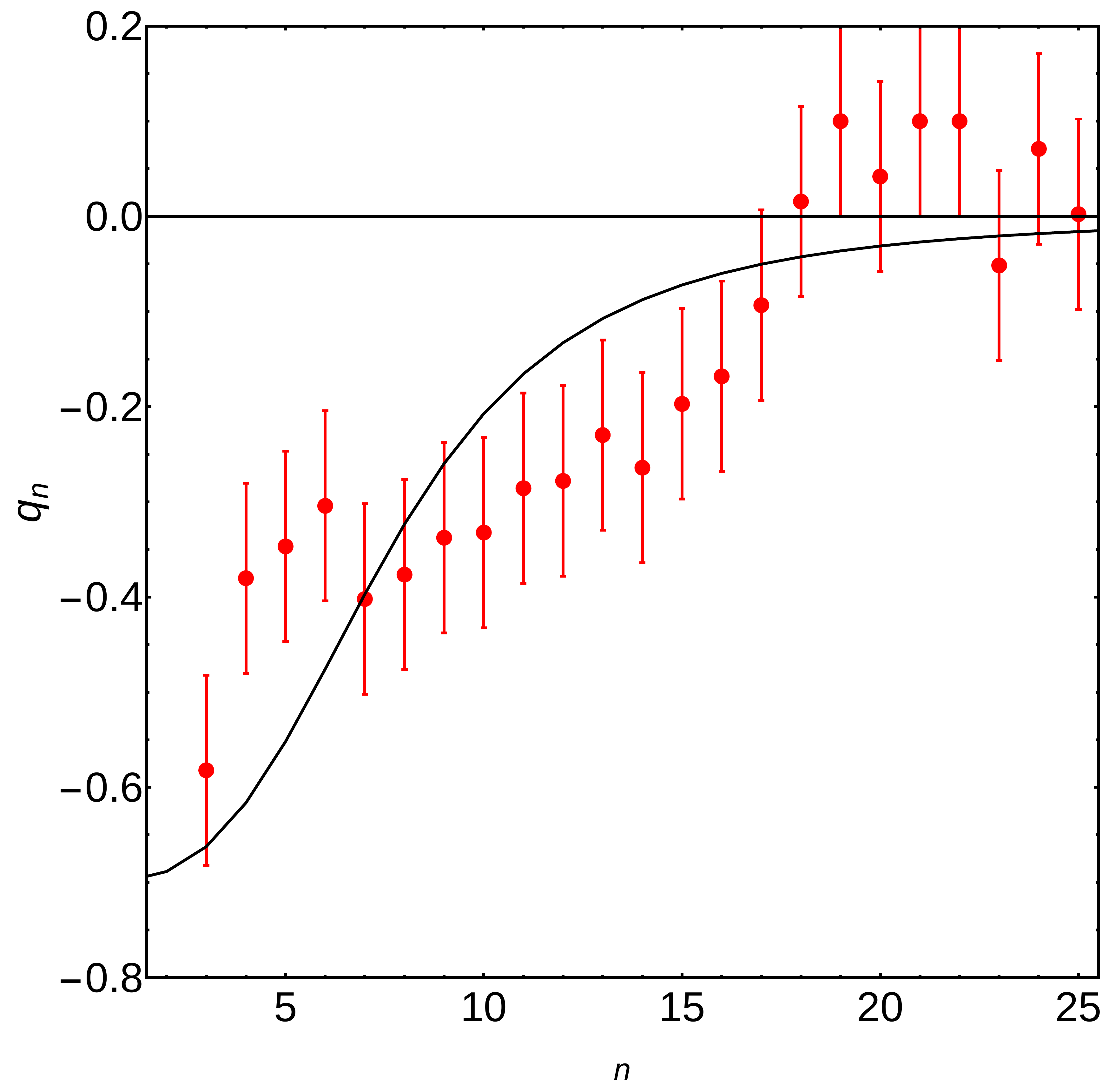}
  \caption{Experimental fitting values for $q_n$ from~\cite{Kazimierczuk2014} (red dots), and theoretical curve according to Eq.~(\ref{eq.qnq}) with $\alpha=7\times10^{-4}$ and $P\cdot q=-0.7$ (black line). The $q_n$ are depicted with an error estimate of $\pm0.1$.}
  \label{fig.qn}
\end{figure}

As we do not yet have a full theory for the peak asymmetry $q_n$, we can use the following argument. 
According to To\-yo\-za\-wa~\cite{Toyozawa64}, the asymmetry comprises couplings to all other excited states via a phonon background.
For large $n$, one may assume that this coupling is rather insensitive as the main couplings to low-$n$ states is almost the same
for similarly large values of $n$. In this case the $q_n$ in the Rabi model would become constant, $q_n\to q$ for $n\to\infty$.
With this in mind, we can set the Toyozawa and the Rabi spectrum equal and determine the apparent $q_n$ fitted in the former.
This yields
\begin{equation}
\begin{split}
	q_n=&g_n^2\Omega_n^2\frac{(\tfrac{\Gamma_n}{2})^2+\delta_n^2}{[(\tfrac{\tilde\Gamma_n}{2})^2+\delta_n^2]^2}q\\
	&+\frac{\Gamma_n}{4\delta_n}\left[1-g_n^2\Omega_n^2\frac{(\tfrac{\Gamma_n}{2})^2+\delta_n^2}{[(\tfrac{\tilde\Gamma_n}{2})^2+\delta_n^2]^2}\right]
\end{split}
\end{equation}
Here, $\tilde\Gamma_n=\sqrt{(\sqrt2-1)(\Gamma_n^2+8\Omega_n^2)}$ in order to include the different fitted widths from the two models. Unsurprisingly, $q_n$ is now a function of $\omega_\text L$, meaning that we can only evaluate it at certain laser frequencies. We thus evaluate the arithmetic average between the values at the two half width values $\delta_n=\pm\tilde\Gamma_n/2$. Including again the different $n$-dependencies, we obtain after some algebra
\begin{equation}
	q_n=P\frac{1+\alpha n^3}{(1+2\alpha n^3)^2}q,\label{eq.qnq}
\end{equation}
with $P$ and $\alpha$ positive constants. For $\alpha n^3\ll1$ we reobtain a constant behaviour of the apperent asymmetries, whereas for large $n$, even though $q$ is constant $q_n$ goes down to zero. This behaviour seems to be roughly confirmed by the experimental fitting to the Toyozawa spectrum, see Fig.~\ref{fig.qn}.

\end{document}